# Learn-Memorize-Recall-Reduce: A Robotic Cloud Computing Paradigm

Shaoshan Liu, Bolin Ding, Jie Tang, Dawei Sun, Zhe Zhang, Grace Tsai, and Jean-Luc Gaudiot

**ABSTRACT**

The rise of robotic applications has led to the generation of a huge volume of unstructured data, whereas the current cloud infrastructure was designed to process limited amounts of structured data. To address this problem, we propose a *learn-memorize-recall-reduce* paradigm for robotic cloud computing. The *learning* stage converts incoming unstructured data into structured data; the *memorization* stage provides effective storage for the massive amount of data; the *recall* stage provides efficient means to retrieve the raw data; while the *reduction* stage provides means to make sense of this massive amount of unstructured data with limited computing resources.

## 1. INTRODUCTION

Robots are mobile devices, and the rise of robotic applications has imposed tremendous pressure on our existing cloud infrastructure. For instance, every day, a mobile phone sends out at least 30 MB of structured data to the clouds of the service provider, the application operators *etc.* (structured data means data that can be easily understood, stored, and retrieved by machines). In contrast, even a very simple robot can easily generate over 1 GB of unstructured multimedia data per day. An extreme form of robot, driverless cars, can generate as much as 2 GB of unstructured data per second [1] (unstructured data means data that cannot be easily understood, stored, and retrieved by machines).

Therefore, we are facing the urgent challenge of designing and implementing a cloud architecture to process this massive robotic unstructured data. For instance, in a real-world scenario, in-home service robots may act in a surveillance role, patrolling the users' house and recording captured videos. Then, users demand the capability of video playback based on intelligent queries using time, location, as well as objects in scene as inputs.

In sections 2, 3, and 4, we first present our implementation of a robotic cloud infrastructure and discuss how this infrastructure can meet this challenge. Specifically, we generalize the requirements into a robotic cloud computing paradigm - *learn-memorize-recall*, where *learning* is about how to automatically understand the unstructured data and convert it into structured data; *memorization* is about how to effectively store the massive amount of data, while *recall* is about how to efficiently retrieve the data when needed.

Nonetheless, as the volumes of data collected from robotic devices exponentially grow over time, having a great robotic cloud infrastructure is not sufficient. Then in section 5, we delve into the final part of the proposed paradigm, *reduction*, the utilization of data reduction techniques to making sense of massive amounts of unstructured data with a limited budget of computing resources.

## 2. ROBOTIC CLOUD ARCHITECTURE

To provide the *learn-memorize-recall* features, we have designed and implemented a cloud architecture, which is illustrated in Figure 1 below. The architecture consists of the following components:

- *Robotic Client Devices*: these devices capture multimedia feeds and send the feeds to the cloud along with their meta data.

- *Streaming Server*: this server handles incoming multimedia and streams on-demand live multimedia feeds to users as requested.

- *Object Recognition*: this is a deep-learning evaluation engine for automatic extraction of semantic information from incoming videos.

- *Key-Value Store*: this key-value store organizes the video feeds along with the learned/extracted semantic information.

- *Query Engine*: this query engine supports retrieval of video feeds. One can search using any combination of time, location, as well as extracted labels.

- *Business Analytics Engine*: this engine generates high-level statistics of all multimedia data. For example, one can be interested in knowing which are the most common objects appear in living rooms, *etc*.

- *Storage Layer*: the storage layer needs to provide high throughput for data persistence and low latency for fast retrieval of video feeds. Also, it must manage heterogeneous storage systems including S3, GCS, Swift, HDFS, OSS, GlusterFS, and NFS.

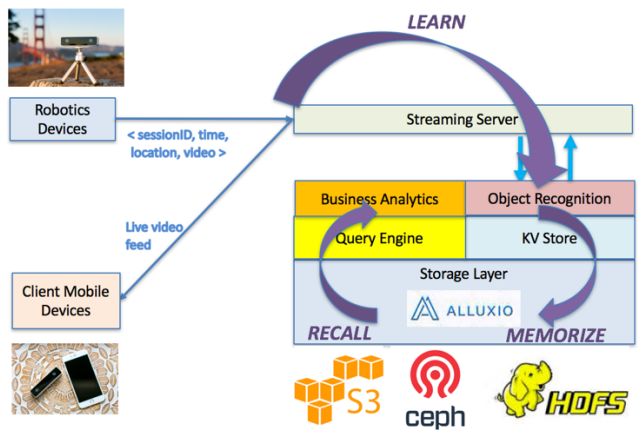

Figure 1: Robotic Cloud Architecture Overview

## 2.1 Learning

As multimedia data comes in to the cloud system, the first task is to learn from the raw data and to extract semantic information out of the multimedia data. For video streams, we can extract object labels from frames and associate these labels with the video stream. For audio streams, we can extract sentences of the spoken language. Then this semantic information can be used as keys, and the raw data streams can be used as values, and together they are stored in the key-value store.

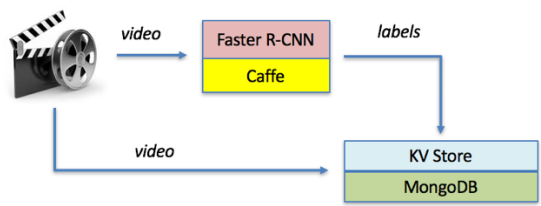

Figure 2: Video Stream Processing

As presented in Figure 2, we need a learning engine that extracts object labels from a video stream. This engine needs to be accurate in terms of recognition rate, and this engine needs to be fast in order for us to capture as many objects as possible in the video. Therefore, in this implementation, we utilize faster r-cnn [3] network running on Caffe [4]. Faster r-cnn introduces a Region Proposal Network (RPN), a fully-convolutional network that simultaneously predicts object bounds and object scores at each position, hence achieving very high detection speed without sacrificing detection accuracy.

## 2.2 Memorization

After extracting the semantic information from the raw multimedia data, the extracted labels, along with the raw data get stored in the key-value store for easy retrieval (Figure 3). We implemented the key-value store using MongoDB [5]. In this case, the key is the meta data including <*sessionID, timestamp, duration, location, <list of labels>>*, and the value is a file path of the raw data in the storage layer.

This poses two challenges: first, in real-world scenarios, users vary in their choice of persistent storage for their data. Some prefer Amazon S3, some their own deployment of HDFS, others Ceph, *etc*. One way to get around this problem is to create a set of APIs for each persistent storage, but this would become impossible to manage as the number of persistent storage options grows. A second, and probably better way, to handle this is to create a unified storage layer to abstract all underlying storages.

To this end, Alluxio enables effective data management across different storage systems through its use of transparent naming and mounting API [6]. Transparent naming maintains an identity between the Alluxio namespace and the underlying storage system namespace. When users create objects in the Alluxio namespace, they can decide whether these objects should be persisted in the underlying storage system. For objects that are persisted, Alluxio preserve the object paths, relative to the underlying storage system directory in which Alluxio objects are stored.

With this feature, we can now manage multiple persistent storages using a single set of storage layer API, which greatly simplifies the management of the memorization part of the robotic cloud computing paradigm.

The second challenge is that the write throughput directly impacts the performance of the whole system. If the write speed of the memorization is slower than the detection speed in learning stage, then it becomes the bottleneck and leads to "memory loss." We will discuss in the next section how we can use Alluxio's tiered storage feature, along with write optimization to improve write throughput.

## 2.3 Recall

The last stage is Recall, in which the users can use any combination of meta data, such as time, location, or detected labels to accurately retrieve the target multimedia data. In our implementation, we use MongoDB to perform the intelligent search using the metadata as keys to first retrieve a file path of the storage layer. Then we issue another request to retrieve the raw data using the file path.

Therefore, the performance of the recall stage greatly depends on the read performance of the storage layer. It is helpful to exploit locality of the storage layer to improve read performance. Ideally, the performance is highest if the requested data is located in the memory of the local machine. However, it is impossible to store all data in the memory of the local machines. We need a mechanism to provide a cache-like structure such that we have different levels of the storage at different speed, including local memory, local Solid State Drives (SSD), local Hard Disk Drives (HDD), and remote storage.

This requirement can be fulfilled by Alluxio's tiered storage feature. With tiered storage, Alluxio can manage multiple storage layers including Memory, SSD, and HDD. Using tiered storage Alluxio can store more data in the system at the same time, since memory capacity may be limited in some deployments. With tiered storage, Alluxio automatically manages blocks between all the configured tiers, so users and administrators do not have to manually manage the locations of the data. Users may specify their own data management strategies by implementing allocators and evictors.

In a way, the Memory layer of the tiered storage serves as the top level cache, SSD serves as the second level cache, HDD serves as the third level cache, while it is the persistent storage is the last level storage. We will discuss in the next section how we can use Alluxio's tiered storage feature, along with prefetching optimization to improve read latency.

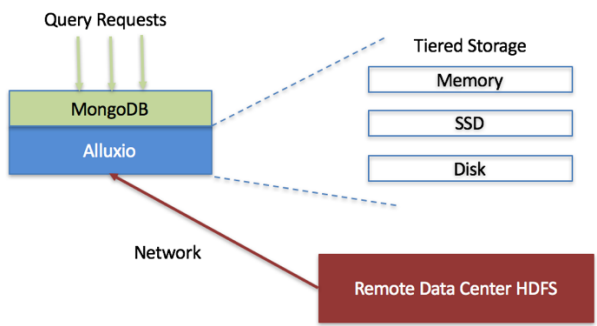

Figure 3: Storage Layer Architecture

## 3. OPTIMIZATIONS

As mentioned in section 2, the storage layer write throughput is critical to the performance of the memorization stage and the storage layer read latency is critical to the performance of the recall stage. In this section, we delve into the optimizations we implement in the storage layer to improve both write throughput and read latency.

### 3.1 Write Optimization

With the default Alluxio tiered storage implementation, when a user writes a new block, it is written to the top tier by default. If there is not enough space for the block in the top tier, the system checks for free space in the next layer. This gets repeated for each layer until free space is found. For instance, if both the memory layer and the SSD layer are full, free space may be found in the HDD layer. Then, the evictor moves a block in the SSD layer to the HDD layer, and then moves a block in the memory layer to the SSD layer. Finally the block can be written back to the memory layer.

This approach is inefficient when the tiered storage is fully occupied as each time a new block is written, the evictor needs to move blocks across all layers before the new block can be written. This greatly reduces the write throughput as it takes a long time to write each block in the storage layer.

To allay this problem, we implement a Direct-Write allocator, such that it writes a block to the first layer that has free space. When we run a stress test to continually write blocks into Alluxio, compared to the default allocator, our Direct-Write approach reduces the write latency by half, and consequently doubling the write throughput.

One could argue that the Write-Direct approach impacts read performance since the new block is now not in the top tier. However, in our experiences, it is very rarely that a block gets read right after being written to storage. Also, if the block gets read, say ten minutes later, by that time, it is highly likely that the block has already been evicted to lower level of the storage layer.

### 3.2 Read Optimization

To optimize read performance, we strive to keep as much data in the frontend servers as possible, thus avoiding the latency of fetching data blocks from remote backend storage servers. We have collected over six months of text-based query data from over 1,000 users. We discovered that, without any optimization, we reached a hit rate of about 50%, meaning that 50% of the queries are satisfied the the frontend servers without hitting the remote backend storage servers, while the other 50% end up hitting the remote servers, thus leading to high latency.

An effective approach to improve this situation is to use prefetching [2]. A simple improvement, when the frontend servers are less busy, we prefetch data from the most requested table into the frontend servers. This simple improvement immediately boosts the hit rate to 70%. For the next step, we break the day into 6 time periods, each is four-hour long. For each time period, we prefetch the most requested table from the same period in the previous day. By doing this, we boost the hit rate to 80%.

This verifies that prefetching is indeed a very effective technique in improving read latency. As we get more queries for multimedia data in our storage, we plan to design more fine-grained prefetching techniques to further improve the performance of read operations. Some of the prefetching strategies we plan to use include:

- Label: since we now have label information associated with each video stream, and we observe an initial trend that people tend to search for certain labels, such as *dog*. Prefetching the videos with the "hottest" label will be an effective way to improve read performance.
- Location: also, now that each video stream is associated with location information, such as bedroom, living room, *etc.*, it will also be effective to prefetch the most searched locations into the frontend storage servers. We will explore these options in the next step.

## 4. PERFORMANCE AND SCALABILITY

In this section we delve into the details of the performance and scalability of the implementation of the learn-memorize-recall robotic cloud computing paradigm. The machine configuration used in this implementation consists of an Intel Core-i7 CPU running at 3 GHz, and a Titan X GPU with 12 GB of graphics memory, and 32 GB of system memory.

## 4.1 Object Recognition Performance

First we look at the performance of the object recognition engine. We stress the engine by feeding it video streams and measure its performance. As shown in Figure 4, when under stress, on a single server, it takes an average of 0.16 sec to process an image. Also, this workload is GPU-bound, as it uses 95% of GPU resources, but only 2~3% of CPU resources and 3% of system memory.

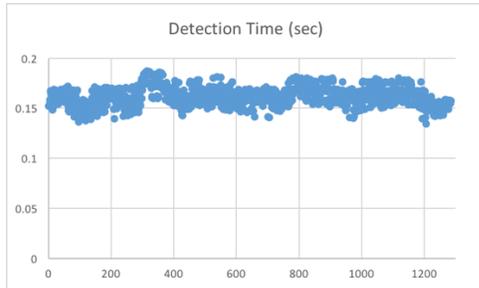

Figure 4: Faster RCNN Detection Stress Test

In our real-world use case, the in-home service robots move at a speed of about 30 centimeters per second. Therefore, in most cases, we only need to extract labels from an image every two seconds without missing an object in the scene. This implies that with each server, we could support 10 simultaneous incoming video streams. This can be further tuned based on the robot speed and user requirements.

## 4.2 Query Engine Performance

We now examine the performance of the query engine. We stress the engine by launching 100 clients repeated send queries to the query engine and we measure the response time. As shown in Figure 5, when under stress, on a single server, it takes on average of 4 ms to process a query. This workload is CPU-bound, in that it uses 98% of CPU resources, none of the GPU resources, and 2% of system memory. This confirms that with one server, we could easily handle over 100 users simultaneously.

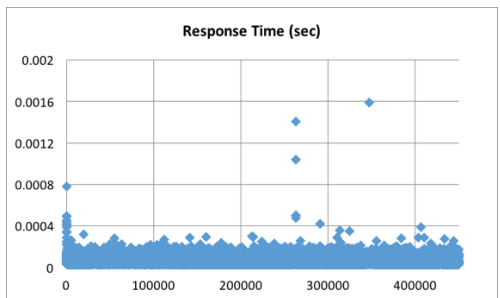

Figure 5: MongoDB Query Stress Test

## 4.3 Storage Performance

Then we examine the performance of the storage engine. We first compare the throughput of a copy operation with Alluxio, with the Native File System, and with Remote HDFS. This parameter is critical as it determines how fast we can write a video feed to storage. If the throughput is too low, then the storage layer may become the bottleneck of the whole multimedia data pipeline. As shown in Figure 6, with Alluxio's in-memory storage engine, we could easily achieve greater than 650 MB/s throughput whereas with Native File System, we could only achieve 120 MB/s. Using remote HDFS, the performance is the worst, only sustaining less than 20 MB/s throughput. This result indicates that for the storage engine not to become the bottleneck of the memorization part, it is important to keep most of the write operations to hit the in-memory storage layer.

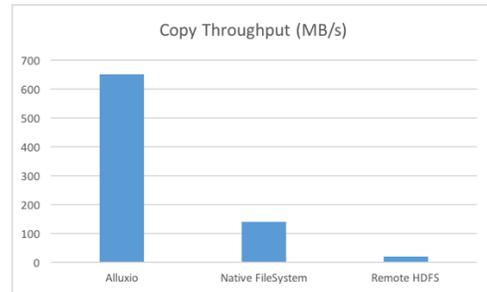

Figure 6: Alluxio Stress Test

Then we evaluate the video retrieval latency, using Alluxio's in-memory storage engine, we are able to retrieve a video within 500 ms. However, when the video is stored in remote machines, the latency can be as high as 20 seconds. Therefore, using Alluxio to buffer "hot" video data could reduce retrieval latencies by as much as 40 folds, which is quite critical to the user experience, especially in the recall stage.

## 4.4 Deployment and Scalability

After understanding the performance of the individual components, we take a look at the deployment setup as well as the scalability of this architecture.

Table 1: Resource Utilization

|  | CPU% | GPU% | MEM% |
| --- | --- | --- | --- |
| Faster RCNN | 3 | 95 | 3 |
| MongoDB Stress | 98 | 0 | 2 |
| Alluxio Stress | 3 | 0 | 95 |

First, Table 1 summarizes the resource utilization of the three components under stress. It is interesting that each component stresses one type of system resource: the faster R-CNN recognition engine stresses the GPU, MongoDB stresses the CPU, while Alluxio stresses the system memory. This indicates that it is best to co-locate these three services on the same servers. This approach provides the following benefits:

- *Cost Efficiency*: as we co-locate the services, we reduce costs.
- *Performance*: the storage layer, Alluxio, is now co-located with the same servers as the learning engine and the query engine, which ensures high write throughput and low read latency.

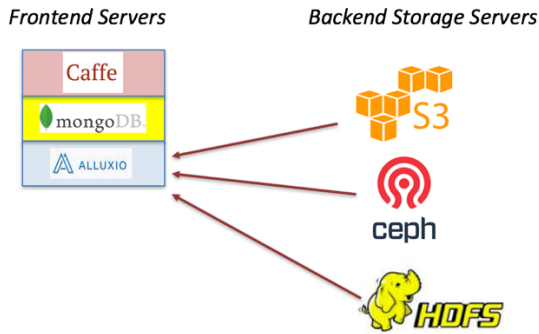

Figure 7: Deployment Architecture

The deployment architecture is shown in Figure 7. It is divided into two parts, the frontend servers and the backend storage servers. The frontend servers process the incoming data and write the raw data in Alluxio. Alluxio then asynchronously persists the raw data in the backend storage servers. Also, when the users generate new queries, the frontend servers process the queries and retrieve the raw data from Alluxio. If the requested data is buffered in Alluxio's tiered storage, it is immediately returned to the user. Otherwise it is a cache miss and Alluxio requests the data from the backend storage servers.

For Alluxio, we use a two-level deployment, allocating 20 GB of memory in the top tier and 200 GB of HDD in the second tiered. With a 10-machine deployment, we provide 2.2 TB of cache space. Assuming 10 MB average video file size, the system can buffer around 200,000 video files. Also, with ten machines, we can simultaneously support 100 video streams as well as 1,000 simultaneous queries. With this setup, we are able to handle 10,000 users.

Next, we seek to answer the question as to whether we could support one million users. The key to this question is whether we could have a 1,000-machine Alluxio deployment running stably. We conducted a stress test, with a 1,000-instance Alluxio deployment and a background script to repeatedly write to and read from the system. After running the script for two weeks, we were able to confirm that Alluxio could reliably scale to 1,000 instances. This confirms that we can use Alluxio to support one million users as our implementation scales.

## 5. NEXT STEP: DATA REDUCTION

As mentioned in the introduction, the volumes of data collected from robotic devices can easily grow exponentially over time, which on one hand, enables our robotic cloud to provide data-analytics and decision-support services to users, but on the other hand, imposes performance challenges on all the three stages, learn-memorize-recall, in terms of both storage and query response time. Although the architecture presented in this paper is optimized for robotic workloads, with a fixed and limited budget of computing resources, the users eventually have to discard more and more data and tolerate longer response time.

### 5.1 Data Reduction Techniques

Data reduction techniques, e.g., sampling, can be a feasible solution to handle the above issue on robotic clouds, especially when small errors are tolerable. For example, when the data collected from robots is used to estimate traffic volume/speed or to learn trajectory patterns, small errors in the results are usually inevitable anyway due to the noise in the initial data collection and in the learning stages. For such tasks, the goal of our robotic cloud is to *provide data-analytics services with low storage cost and interactive query response time at the cost of small errors in the analytical results and answers*.

To achieve this, we utilize techniques from approximate query processing (AQP), which has been studied for decades mostly in the context of relational databases. The goal of AQP is to provide approximate answers to a subclass of SQL queries with interactive response time and estimated error bounds. This line of work has become very active recently because of the explosion of data. The recent efforts result in both innovative techniques and more mature sampling-based systems, e.g., BlinkDB [13], QuickR [11], and Sample+Seek [10].

Interested readers can refer to [12] for a brief survey of the progresses. Sampling-based AQP techniques can be characterized by where the sampler (for the data-reduction purpose) is in the whole pipeline. In one way, the sampler can be placed right before the data is streamed into the query execution engine, or even pushed to the nodes of a query plan tree with the execution engine (e.g., QuickR). In the other way, samplers are used pre-computed samples of the data tables, and SQL queries are rewritten to be executed on these sample tables for fast response (e.g., BlinkDB and Sample+Seek).

### 5.2 Usages in Robotic Clouds

In a robotic cloud, we have more options about where to place the samplers and more considerations to be addressed for each option. Following are our preliminary proposals.

- *Sampling before learning*. We can reduce the amount of raw data collected from users by placing samplers in the client devices or the streaming server. A tuple (e.g., an image) in the raw data is selected into the object recognition component only with some probability, which is determined by some light-weight calculation based on the tuple's meta data, e.g., time and location.
- *Sampling before memorization*. After the semantics information is extracted from the raw data tuples, we can pre-compute and store only a subsample into the key-value store. For example, a row <sessionID, timestamp, duration, location, <list of labels>> is selected into the store with a probability determined by the "list of labels"—we can select more sample rows for some hot and more important labels and less sample rows for the others.
- *Online sampling during recall*. When a query is issued in the recall stage, we can invoke a sampler to select

only a subset of rows from the key-value store into the query execution engine (and of course, rewrite the query to rescale the answer). For example, we may want to count the number of times a label appears between a time interval. A number of sampler can be borrowed from AQP systems [12] to provide estimates of answers and error bounds for such queries.

Note that we can either adopt all the three proposals at the same time or only some of them, based on the different types of underlying tasks and the amount of resource budget.

Finally, it is important to note that we always want to guarantee that the errors are no more than some threshold in our analytical results or answers, or at least, we want to provide estimates of the errors. Such guarantees or estimations are relatively easier for SQL-like aggregation queries (e.g., counting the number of cars on a street), but become more challenging for complex tasks (e.g., predicting the trajectory of a car). We will start with adopting the sampling-based data reduction techniques in our robotic cloud to provide cheap and fast data-analytics services for users with relatively simpler analytical workloads, and extend them for more types of tasks in future.

## 6. CONCLUSIONS

We are facing the data explosion problem from robotic applications. Unlike existing mobile devices, robots generate massive amount of unstructured data, which can not be easily understood, stored, and retrieved by machines.

To address this problem, in this paper, we proposed a *learn-memorize-recall-reduce* paradigm for robotic clouds: the *learning* stage extracts semantic information from incoming unstructured data and convert it into structured data; the *memorization* stage provides effective storage for the massive amount of data; the *recall* stage provides efficient means to retrieve the raw data; and the *reduction* stage provides means to making sense of massive amount of unstructured data with limited computing resources.

Dr. Shaoshan Liu is currently the Co-Founder of PerceptIn, working on developing the next-generation robotics platform. Contact him at: shaoshan.liu@perceptin.io

Dr. Bolin Ding is currently a Researcher in the Data Management, Exploration and Mining group at Microsoft Research, Redmond. Contact him at: bolin.ding@microsoft.com

Dr. Jie Tang is the corresponding author and she is currently an associate professor in the School of Computer Science and Engineering of South China University of Technology, Guangzhou, China. Contact her at: cstangjie@scut.edu.cn

Dawei Sun is currently with Tsinghua University and PerceptIn, working on Deep Learning and cloud infrastructures and autonomous robots. Contact him at: sdw14@mails.tsinghua.edu.cn

Dr. Zhe Zhang is the co-founder of PerceptIn, working on developing the next-generation robotics platform. Contact him at: zhe.zhang@perceptin.io

Dr. Grace Tsai is a founding engineer of PerceptIn, working on developing the next-generation robotics platform. Contact her at: grace.tsai@perceptin.io

Dr. Jean-Luc Gaudiot is professor in the Electrical Engineering and Computer Science Department at the University of California, Irvine and is currently serving as the 2017 President of the IEEE Computer Society. Contact him at gaudiot@uci.edu